\documentclass[aps,prb,twocolumn,groupedaddress]{revtex4}
\usepackage{amsmath}
\usepackage{amssymb}
\usepackage{graphicx}
\usepackage{color}
\usepackage{dcolumn}
\begin{document}

\title{Landauer-B\"uttiker Study of the Anomalous Hall Effect}

\author{Maria Silvia Garelli and John Schliemann}

\address{Institute for Theoretical Physics, University of
Regensburg, D-93040 Regensburg, Germany}

\date{\today}

\begin{abstract}
We report on Landauer-B\"uttiker studies of anomalous Hall transport
in a two-dimensional electron gas with Rashba spin-orbit coupling and a
magnetization provided by localized magnetic moments.
Our system is
described by a discretized tight-binding model in a four-terminal geometry.
We consider both the case of magnetically disordered systems as well as
ballistic transport in disorder-free systems
with spatially homogeneous magnetization. In the latter case we investigate
both out-of-plane and in-plane magnetizations. We numerically
establish a close connection between singularities in the density of states
and peaks in the Hall conductance close to the lower band edge. Consistent with
previous theoretical studies based on diagrammatic perturbation
expansions, these peaks occur at Fermi energies where only the
lower dispersion branch is occupied. Moreover, for large magnetization
the Hall conductance is, along with the density of states, suppressed.
This numerical finding can be understood from analytical properties of
the underlying model in the limit of an infinite system.
\end{abstract}


\maketitle

\section{Introduction}
\label{intro}

The anomalous Hall effect (AHE) is the subject of a long-standing and
partially still ongoing theoretical debate
\cite{Karplus54,Smit55,Berger70,Nozieres73,Jungwirth02,Sinova04a,Sinitsyn08}.
It amounts in a Hall conductivity which is not due to an external magnetic
field but the result of the magnetization of a solid.
A large portion of the renewed interest in this phenomenon is generated by
research on ferromagnetic semiconductors \cite{MacDonald05,Jungwirth06}.
In general, it is common to distinguish between two types of mechanisms
for the AHE, both relying on spin-orbit interaction: The extrinsic mechanism
requires the presence of impurities or other imperfections and is
based on contributions to spin-orbit coupling from such scattering potentials.
This spin dependence of the effective scattering potential gives rise to the
skew-scattering \cite{Smit55} and the side-jump
\cite{Berger70} contribution to the anomalous Hall conductivity.
The intrinsic mechanism is independent of scattering centers and is a result
of the spin-orbit-coupled electronic band structure, where the
spin-orbit interaction stems from the ordered crystal lattice itself.

Among many different systems, the case of a two-dimensional
semiconductor electron gas with an intrinsic effective spin-orbit
coupling of the Rashba type \cite{Rashba60} has attracted considerable
theoretical interest \cite{Culcer03,Dugaev05,Sinitsyn05,Liu05,Liu06,Onoda06,Inoue06,Kato07,Borunda07,Nunner07,Nunner08,Kovalev08,Kovalev09} and was also studied experimentally in an $n$-doped II-VI semiconductor heterostructure
containing manganese ions
\cite{Cumings06}. The theoretical investigations have considered both the
intrinsic effect \cite{Inoue06,Kato07} as well as
combinations of intrinsic and extrinsic mechanisms
\cite{Culcer03,Dugaev05,Sinitsyn05,Liu05,Liu06,Onoda06,Borunda07,Nunner07,Nunner08,Kovalev08,Kovalev09}.
In the present work we shall concentrate on the purely intrinsic AHE.

An important tool for the theoretical description
of transport in such mesoscopic systems is given by the Landauer-B\"uttiker
formalism \cite{datta,ferry}.
In this paper we report on numerical studies within this approach
on Hall transport in a two-dimensional electron gas with Rashba spin-orbit
interaction and magnetic impurities. Such an investigation of the AHE seems
to be missing in the previous literature although several studies of this kind
on the related spin Hall effect are available
\cite{Hankiewicz04,ShengL05,LiJ05a,Nikolic05a,Nikolic05b,Hankiewicz05,Erlingsson05,Moca05,Nikolic06}.

This paper is organized as follows. In section \ref{model} we describe
important properties of our underlying model and outline the
Landauer-B\"uttiker approach to transport in such systems. More specific
information on Green\rq s function in semi-infinite leads used in our
study can be found in appendix \ref{greenfunct}. In section \ref{results}
we present our numerical results covering both ballistic transport and
transport in magnetically disordered systems. We close with conclusions and
an outlook in section \ref{concl}.

\section{Model and Approach}
\label{model}

We consider a two-dimensional gas of non-interacting electrons with
spin-orbit interaction of the Rashba type \cite{Rashba60}. Additionally,
the electron spin is coupled to magnetic impurities.

\subsection{Continuum model}
\label{continuum}
The generic
single-particle Hamiltonian for the continuum system reads
\begin{equation}
{\cal H}=\frac{\vec p^{2}}{2m^{\ast}}+\frac{\alpha}{\hbar}
\left(p_{x}\sigma^{y}-p_{y}\sigma^{x}\right)+\vec\Delta\cdot\vec\sigma\,.
\label{contham}
\end{equation}
Here $\vec p$ is the electron momentum, $m^{\ast}$ its effective band mass, and
the Pauli matrices $\vec\sigma$ describe the electron spin.
The strength of the spin-orbit
interaction is described by the Rashba parameter $\alpha$, and $\vec\Delta$
is the effective Zeeman splitting due to the coupling of the
electron spin to magnetic impurities. In general, this quantity will
be position-dependent, $\vec\Delta=\vec\Delta(\vec r)$. However, it is also
instructive to consider the case of spatially constant
magnetization corresponding to an uniform impurity polarization.
In the following we will consider both the case of a homogeneous
magnetization perpendicular to the
plane of the electron gas, as well systems with
in-plane magnetization. We note that an in-plane magnetization
can also be interpreted as a genuine magnetic field which couples
in a strictly two-dimensional situation only to the spin of the electron
but not to its orbital degrees of freedom.
For a homogeneous magnetization the eigenstates are given by
plane waves and the energy of a given wave vector
$\vec k=\vec p/ \hbar$ reads
\begin{eqnarray}
\varepsilon(\vec k) & = & \frac{\hbar^{2}k^{2}}{2m^{\ast}}\nonumber\\
& & \pm
\sqrt{(-\alpha k_{y}+\Delta_{x})^{2}+(\alpha k_{x}+\Delta_{y})^{2}+\Delta_{z}^{2}}\,.
\label{contdis}
\end{eqnarray}
Let us first discuss the case of a purely perpendicular
magnetization, $\vec\Delta=(0,0,\Delta)$. Here one finds that,
provided that the energy scale of the Rashba coupling
$\varepsilon_{R}:=m^{\ast}\alpha^{2}/ \hbar^{2}$ is larger than the Zeeman coupling,
\begin{equation}
\varepsilon_{R}>|\Delta|\,,
\label{contcond}
\end{equation}
the lower dispersion branch in Eq.~(\ref{contdis}) has a minimum at finite
$k=k_{min}$,
\begin{equation}
k_{min}=\frac{1}{|\alpha|}\sqrt{\varepsilon_{R}^{2}-\Delta^{2}}
\end{equation}
with minimum energy
\begin{equation}
\varepsilon_{min}=\varepsilon_{-}(k_{min})=-\frac{1}{2}\varepsilon_{R}
-\frac{1}{2}\frac{\Delta^{2}}{\varepsilon_{R}}\,.
\end{equation}
This dispersion minimum at finite wave vector leads to a van Hove singularity
in the electronic density of states at the bottom of the lower branch,
$\varepsilon\to\varepsilon_{min}+0$. Explicitly, the density of states is given
by
\begin{equation}
D(\varepsilon)=\left\{
\begin{array}{cc}
0 & \varepsilon<\varepsilon_{min}\\
\frac{m^{\ast}}{\pi\hbar^{2}}
\sqrt{\frac{\varepsilon_{R}}{2(\varepsilon-\varepsilon_{min})}}
& \varepsilon_{min}\leq\varepsilon<-|\Delta|\\
\frac{m^{\ast}}{2\pi\hbar^{2}}\left(1+
\sqrt{\frac{\varepsilon_{R}}{2(\varepsilon-\varepsilon_{min})}}\right)
& -|\Delta|\leq\varepsilon<|\Delta|\\
\frac{m^{\ast}}{\pi\hbar^{2}} & |\Delta|\leq\varepsilon
\end{array}
\right.
\label{vanHove}
\end{equation}
This quantity has obviously a square-root singularity at
$\varepsilon\to\varepsilon_{min}+0$.
As we shall see below, such singularities
are intimately linked to the observation of anomalous Hall transport.
Note also that $D(\varepsilon)$ is discontinuous
(but finite) at $\varepsilon=\pm|\Delta|$.
These discontinuities vanish for $\Delta=0$.
On the other hand, if the Zeeman coupling dominates over the spin-orbit
interaction,
\begin{equation}
\varepsilon_{R}<|\Delta|\,,
\end{equation}
the energy dispersion branches have stationary points only at zero
wave vector. Here no van Hove singularity occurs in the density of states, apart from a step-like behaviour at minimum energy.
\begin{figure}
   \centering
   \includegraphics[height=6cm,keepaspectratio=true]{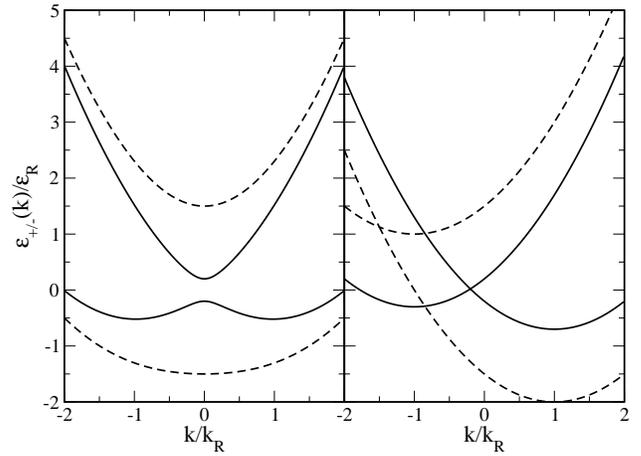}
   \caption{Energy dispersion $\varepsilon_{\pm}(\vec k)$ for
$\Delta=0.2\varepsilon_{R}$ (solid lines) and $\Delta=1.5\varepsilon_{R}$
(interrupted lines).
Left panel: perpendicular magnetization
$\vec\Delta=(0,0,\Delta)$. Right panel: in-plane magnetization
$\vec\Delta=(\Delta_{x},\Delta_{y},0)$ with the wave vector $\vec k$
being orthogonal to $\vec\Delta$, $\vec\Delta\cdot\vec k=0$.
The energies are given in units of the Rashba energy $\varepsilon_{R}$ while
the wave vector is measured in units of the inverse ``Rashba wave length''
$k_{R}=m^{\ast}\alpha/ \hbar^{2}$.}
\label{dispersions}
\end{figure}
These two cases $\varepsilon_{R}\gtrless\|\Delta|$ are illustrated in the
left panel of Fig.~\ref{dispersions}. Note that for $\varepsilon_{R}>|\Delta| $ the avoided crossing at $k=0$ does not lead to a van Hove singularity since these stationary points occur at vanishing wave vector.

In the case of an arbitrary magnetization direction,
$\vec\Delta=(\Delta_{x},\Delta_{y},\Delta_{z})$, a closer analysis shows
that the stationary points of the dispersion (fulfilling
$\partial\varepsilon_{\pm}/ \partial\vec k=0$) lie at wave vectors perpendicular
to the in-plane projection of the magnetization,
\begin{equation}
\vec k\cdot\vec\Delta=0\,.
\end{equation}
However, since this very general case does not seem to allow for further
explicit results, we shall concentrate on a purely in-plane magnetization,
$\vec\Delta=(\Delta_{x},\Delta_{y},0)$.
Here the stationary points of the dispersion correspond to energy
minima and lie at wave vectors
\begin{equation}
\vec k_{min}^{\pm}=\pm\frac{m^{\ast}\alpha}{\hbar^{2}}
\frac{(-\Delta_{y},\Delta_{x})}{\Delta}\,.
\end{equation}
The absolute minimum is given by
\begin{equation}
\varepsilon_{min}^{-}=\varepsilon_{-}(\vec k_{min}^{-})=-\frac{1}{2}\varepsilon_{R}
-\Delta
\end{equation}
(assuming $\alpha>0$),
while another minimum occurs at $\vec k=\vec k_{min}^{+}$ with
\begin{equation}
\varepsilon_{min}^{+}=\varepsilon_{\pm}(\vec k_{min}^{+})=-\frac{1}{2}\varepsilon_{R}
+\Delta\,,
\end{equation}
where the plus (minus) sign in $\varepsilon_{\pm}(\cdot)$ applies if
$\Delta>\varepsilon_{R}$ ($\Delta<\varepsilon_{R}$).
These two cases  are depicted   in the right panel of Fig.~\ref{dispersions}.
The dispersion minima at finite wave
vector $\vec k=\vec k_{min}^{\pm}$ are physically easily understood: In the
case $\varepsilon=\varepsilon_{min}^{-}$, the Zeeman field the
and spin coupling provided by the Rashba interaction are parallel to each other
leading to an energetic minimum for the appropriate spin direction,
while for the higher minimum $\varepsilon=\varepsilon_{min}^{+}$ these
couplings are antiparallel. Note also that these energetic minima remain
at finite wave vector for arbitrarily large magnetization. Therefore,
differently from the case of perpendicular magnetization, the
van Hove singularities in the density of states do not vanish for large
Zeeman coupling.

\subsection{Discrete system}
\label{discrete}
The standard discretized version of the Hamiltonian (\ref{contham})
on a square lattice reads
\begin{eqnarray}
{\cal H} & = & -t\sum_{m,n;\alpha}[c^{\dagger}_{m,n;\alpha}c_{m+1,n;\alpha}\nonumber\\
 & & \qquad\qquad+c^{\dagger}_{m,n;\alpha}c_{m,n+1;\alpha}+{\rm h.c.}]\nonumber\\
 & & +\lambda\sum_{m,n;\alpha,\beta}
[-ic^{\dagger}_{m,n;\alpha}\sigma^x_{\alpha,\beta}c_{m,n+1;\beta}
\nonumber\\
 & & \qquad\qquad
+ic^{\dagger}_{m,n;\alpha}\sigma^y_{\alpha,\beta}c_{m+1,n;\beta}+{\rm h.c}]\nonumber\\
 & & +\sum_{m,n;\alpha,\beta} c^{\dagger}_{m,n;\alpha}
\vec\Delta_{m,n}\cdot\vec\sigma_{\alpha,\beta}c_{m,n;\beta}\,.
\label{discham}
\end{eqnarray}
Here $m$ and $n$ label lattice locations with respect to the $x$- and $y$-axis,
respectively, and $\alpha$,$\beta$ are spin indices. The hopping parameter
$t$ is related to the effective mass $m^{\ast}$ and the lattice constant $a$
via $t=\hbar^{2}/2m^{\ast}a^{2}$, and the parameter $\lambda$ is given by
$\lambda=\alpha/2a$. To give a specific example, for a host material like
gallium arsenide we have an effective band mass of $m^{\ast}=0.067m_{0}$ (with
$m_{0}$ being the free electron mass) and a lattice spacing of
$a=5.6\AA$ leading to a hopping parameter $t=1.8{\rm eV}$. Typical values
for the Rashba parameter are of order $0.1{\rm eV}\AA$ such that we
have typically $\lambda\approx 0.01t$.

For a spatially homogeneous
impurity polarization of arbitrary direction,
$\vec\Delta_{m,n}=(\Delta{x},\Delta{y},\Delta_{z})$
the energy dispersions are given by
\begin{eqnarray}
 \varepsilon_{\pm}(\vec k) & = & -2 t (\cos(k_x a)+\cos(k_y a))\nonumber\\
 & \pm & \Biggl[(\Delta_{x}-2\lambda\sin(k_y a))^{2}\nonumber\\
 & & +(\Delta_{y}+2\lambda\sin(k_x a))^{2}+\Delta_{z}^{2}\Biggr]^{1/2}\,.
\label{discdis}
\end{eqnarray}
In order to make contact to the continuum model, one has to evaluate
this dispersion for small wave vector, $ka\ll 1$, corresponding to the lower
band edge, where it reproduces the Eq.~(\ref{contdis}) up to a rigid shift of
$(-4t)$ which is just half of the band width in the absence of magnetization.

However, these dispersion relations (\ref{discdis}) of the discrete system
lead to very intricate conditions for
stationary points which do not seem to be explicitly solvable. We therefore
concentrate on the case of a homogeneous polarization perpendicular
to the plane, $\vec\Delta_{m,n}=(0,0,\Delta)$.
Here the lower branch leads again to a singularity in the
density of states provided that
\begin{equation}
2\lambda^{2}>t|\Delta|\,,
\end{equation}
which is exactly the same as the condition (\ref{contcond}).

\subsection{ Hall bridge and Landauer-B{\"u}ttiker formalism}
\label{LB}
\begin{figure}
\begin{center}
\input{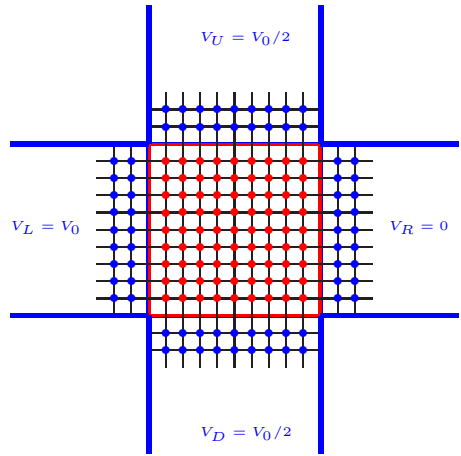}
\end{center}
\caption{Four-terminal Hall bridge. The central region is described by
the discretized Hamiltonian (\ref{discham}) incorporating
Rashba spin-orbit coupling and magnetic impurities.}
\label{2DEG}
\end{figure}

The above discretized system described by the Hamiltonian
(\ref{discham}) is studied as the central region
of a four-terminal Hall bridge shown in Fig.~\ref{2DEG}.
In the following, we will investigate this system consisting of its
central part and ideal semi-infinite leads without spin-orbit coupling and
magnetization using the Landauer-B{\"u}ttiker formalism \cite{datta,ferry}.
We now briefly summarize the most important features of this method
as applied to the calculation of the Hall conductivity.

Within the Landauer-B{\"u}ttiker approach, the Hall conductivity is given by
\begin{equation}\label{CHC}
 \sigma_{H}
=\frac{I_U}{V_L-V_R}=-\frac{1}{2}G_{UL}+\frac{1}{2}G_{UR}\,,
\end{equation}
where $I_U$ is the current flowing into lead up, and $G_{UL}$ and $G_{UR}$
are the conductances between lead up and left and between lead up and
right, respectively. These quantities
can be calculated via the following equation
\begin{equation}
 I_p=\sum_q G_{pq}(V_p-V_q)\,,
\end{equation}
which describes the current flowing in lead $p$, where
\begin{equation}
 G_{pq}=\frac{e^{2}}{h}T_{pq}(\varepsilon_f)
\end{equation}
is the charge conductance between leads $p$ and $q$. This quantity
is proportional to the transmission function $T_{pq}$ defined as
\begin{equation}\label{transmission}
 T_{pq}={\rm tr}[\Gamma_p G^r\Gamma_q G^a]\,,
\end{equation}
where the retarded (advanced) Green\rq s function $G^{r(a)}$ and the
line-width function $\Gamma$ enter. These functions are the heart of the
Landauer-B{\"u}ttiker approach. Indeed, both Greens\rq s functions
\begin{equation}
 G^r=[G^a]^{\dagger}=\left(\varepsilon_f-H-\sum_q\Sigma^r_q\right)^{-1}
\end{equation}
 and
\begin{equation}
 \Gamma_q=i[\Sigma_q^r-\Sigma_q^a]
\end{equation}
depend on the retarded (advanced) self energy $\Sigma^{r(a)}$.
The retarded $\Sigma^r_{q,\mu}$ and the advanced
$\Sigma^{a}_{q,\mu}=\Sigma^{r\dagger}_{q,\mu}$ self-energy terms describe the
coupling between the two-dimensional electron gas (2DEG) in
the central region and the four ideal
semi-infinite leads and can be formulated as
\begin{equation}\label{green}
 \Sigma^r_{q,\mu}(i,j)=t^2 g^r_{q,\mu}(p_i,p_j),
\end{equation}
where $g^r_{q,\mu}$ is the Green\rq s
function of the isolated semi-infinite lead.
The only terms which enter into the self-energy are the coupling terms
between each lead and the 2DEG central region. The coupling between a
lead and the central system give rise to a coupling matrix which is
non-zero only for adjacent points $i$ (lying at the edge of the 2DEG) and
$p_i$ (lying at the lead큦 edge which faces the 2DEG). Since the lead
Green\rq s
function can be evaluated analytically \cite{datta,ferry}, the self-energy
method can be exploited to deal with an infinite system, such as a
semi-infinite lead, by calculating only the Green\rq s
function of a finite region.
The analytical expression of our Green\rq s function is derived in
appendix \ref{greenfunct}.

Another important quantity which can be
calculated within the Landauer-B{\"u}ttiker formalism is the density of
the states (DOS) given by
\begin{equation}
 D(\varepsilon)=\frac{1}{2\pi N^{2}a^{2}}{\rm tr}[A(\varepsilon)]
=-\frac{1}{2\pi N^{2}a^{2}}{\rm tr}[{\rm Im}(G^r)],
\label{DOS}
\end{equation}
where $A(\varepsilon)=i[G^r-G^a]$ is the spectral function,
and $N^{2}$ is the number of lattice sites in the central region taken to be
a square. The above expression is identical to another
standard textbook result,
\begin{eqnarray}
D(\varepsilon) & = &  \frac{1}{(2\pi)^2}\int d^2 \vec k
\delta(\varepsilon-\varepsilon(\vec k))\\
& = & \frac{1}{(2\pi)^2}\oint_{\varepsilon(\vec k)=\varepsilon}dk
\frac{1}{|\nabla_{\vec k}\varepsilon|}\,,
\end{eqnarray}
which can be used to derive , e.g., Eq.(\ref{vanHove}).
For an infinite 2DEG with dispersion
$\varepsilon(\vec k)=-2 t(\cos(k_x a)+\cos(k_y a))$ the
DOS can be calculated analytically leading to a logarithmic divergence at
$\varepsilon=0$ and a saturation to constant values at the edges of the band,
i.e. $D(\varepsilon=\pm 4 t)=\frac{1}{4 \pi}t a^2$.
On the other hand,
in the case of a finite 2DEG central region and no coupling to the leads, the
DOS is just given by a sum of $\delta -$peaks for values of $\varepsilon_f$
which match the eigenvalues of the Hamiltonian of the central conductor.
As we shall see below, the anomalous Hall conductance is closely related to the
DOS.

\section{Results}
\label{results}

Let us now describe our numerical results based on the Landauer-B\"uttiker
formalism outlined before. We first concentrate on disorder-free ballistic
systems.

\subsection{Ballistic Hall Transport}
\label{ballistic}
Here we present our results for a disorder-free central region with
homogeneous magnetization $\vec\Delta$. We will both cover the case
of magnetization perpendicular to the plane of the 2DEG,
$\vec\Delta=(0,0,\Delta)$ and the case of in-plane magnetization
of various directions. In the latter scenario, the magnetization
$\vec\Delta$ can also be interpreted as a proper magnetic field
$\vec B$ which, in a strictly two-dimensional system, couples only to the
spin but not to the orbital degrees of freedom of charge carriers.

\subsubsection{Magnetization perpendicular to the 2DEG}
\label{perp}
\begin{figure}
\centering
\includegraphics[scale=0.35]{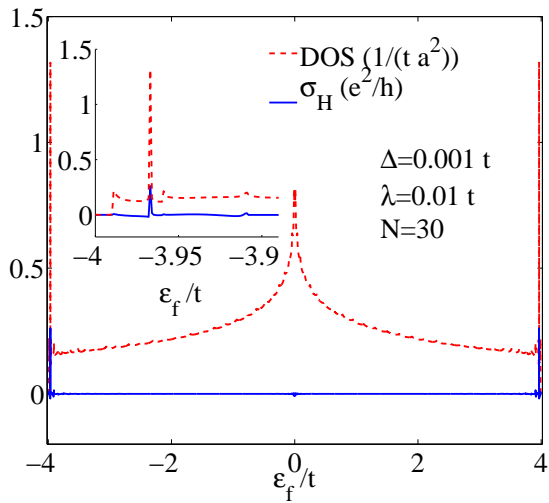}
\caption{(Color online) Hall conductance and DOS
for a Zeeman coupling $\Delta=0.001 t$ and a Rashba parameter$\lambda=0.01 t$
as a function for Fermi energy
$\varepsilon_f\in [-4 t-\Delta, 4 t+\Delta]$ and a linear system size of
$N=30$. The inset shows the behavior of the Hall conductance and the DOS
close to the lower band edge. Both quantities are characterized by a simultaneous
peak.}\label{CHC_total}
\end{figure}

Let us first turn to the case of systems magnetized perpendicularly to the
plane of the 2DEG. We consider a magnetization $\vec\Delta=(0,0,\Delta)$
and have evaluated the Hall conductance and the density of states
according to Eqs.~(\ref{CHC}) and (\ref{DOS}), respectively. As is must be,
the Hall conductance vanishes for zero magnetization since the conductances
$G_{UL}$ and $G_{UR}$ are identical and cancel out.
With a finite Zeeman coupling, however, a charge current $I_U$ starts to
flow in lead UP signalling a finite Hall conductance.
Fig.\ref{CHC_total} shows the Hall conductance along with the DOS
for a Zeeman coupling $\Delta=0.001 t$ and a Rashba parameter of
$\lambda=0.01 t$ as a function of Fermi energy
$\varepsilon_f\in [-4 t-\Delta, 4 t+\Delta]$. Both quantities plotted are
perfectly symmetric with respect to the band center $\varepsilon_f=0$. The Hall
conductance is characterized by an oscillatory behavior over the
entire energy range with particularly dominating peaks near the band edges;
a smaller peak occurs also at the band center. On the other hand, the DOS
shows the predicted logarithmically divergence at $\varepsilon_f=0$ and
in addition even more pronounced singularities near the band edges
at exactly the same positions as the peaks of the Hall conductance.

\begin{figure}
\centering
\includegraphics[scale=0.35]{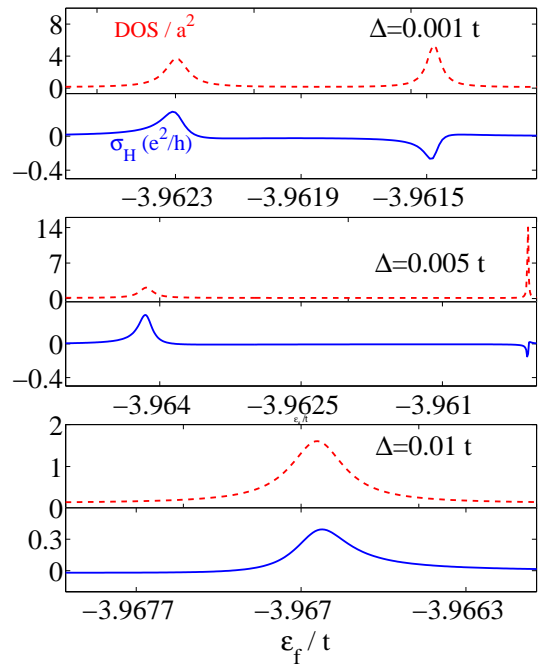}
\caption{(Color online) DOS (red dashed line) in units $1/ta^2$ and
Hall conductance (blue solid line) in units of $e^2/h$
near the lower band edge for the same system size and Rashba coupling as
in Fig.~\ref{CHC_total} but various Zeeman couplings $\Delta$. In all
cases an obvious correspondence between the extrema DOS and Hall
conductance occurs.
}\label{DOS_CHC_comp}
\end{figure}

As we wish to make contact to previous theoretical investigations on
anomalous Hall transport in 2DEGs described by continuum models\cite{Culcer03,Dugaev05,Sinitsyn05,Liu05,Liu06,Onoda06,Inoue06,Kato07,Borunda07,Nunner07,Nunner08,Kovalev08,Kovalev09}, we will
concentrate in the following on the peaks of Hall conductance and DOS
close to the lower band edge.
In Fig.\ref{DOS_CHC_comp} we have plotted both quantities near the
lower band edge for the same
Rashba parameter and system size as before, but different Zeeman couplings
$\Delta$.
As seen in the figure, extrema of the DOS and the Hall conductance
occur at the same position in energy, independently of the regime
of Zeeman coupling.
The close correspondence between extrema of the DOS and the Hall conductance
will be an important finding for our
further analysis of anomalous Hall transport.
Note that the maxima of the DOS at lower energies become weaker with
increasing Zeeman coupling $\Delta$. Such a behavior can indeed be expected
from the analytically accessible properties of the infinite
system discussed in section \ref{model}. Here the divergent van Hove singularities
in the DOS vanish if the magnetization dominates the spin-orbit coupling.
The systems investigated in this work numerically are obviously different as
they are finite and coupled to semi-infinite leads lacking spin-orbit
interaction. However, the above observations shall still guide our
intuition regarding the interplay between magnetization and spin-orbit
coupling.

\begin{figure}
\centering
\includegraphics[scale=0.38]{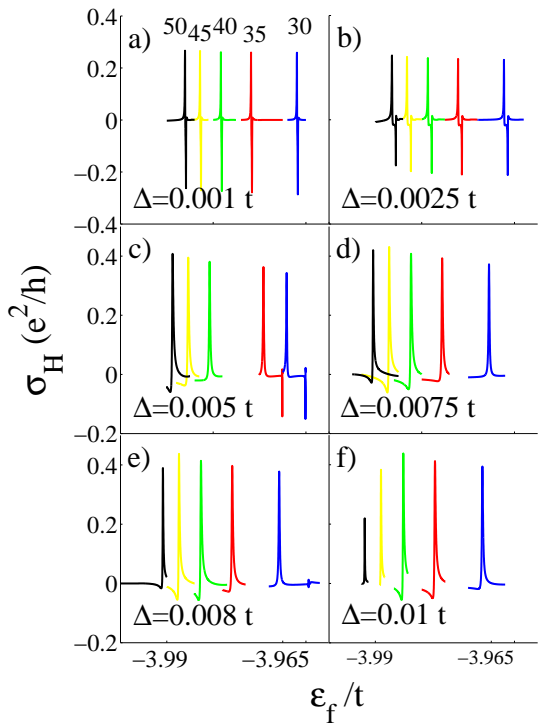}
\caption{(Color online) Hall Conductance near the bottom of the band
for a Rashba parameter of $\lambda=0.01t$ and various Zeeman couplings
$\Delta$. The linear system size varies from $N=30$ (blue, right), $N=35$ (red),
$N=40$ (green), $N=45$ (yellow), and $N=50$ (black, left).
\label{CHC_collectiveB001}}
\end{figure}

Fig.~\ref{CHC_collectiveB001} displays the Hall conductance near the bottom of
the band for a Rashba parameter of $\lambda=0.01t$ and various Zeeman
couplings $\Delta$ with the linear system size varying from $N=30$ to $N=50$.
For small magnetization $\Delta\lesssim 0.001t$
the height of the Hall conductance peaks is approximately independent of
the system size, while for larger Zeeman couplings
$0.001\lesssim\Delta\lesssim 0.0075t$ slightly
grows with increasing system size. For even larger
$\Delta\gtrsim0.0075t$
a decrease is observed for large system sizes.
This qualitative behavior persists in a range of Rashba parameters
 $\lambda\in [0.005 t, 0.02 t]$ with the above threshold values for the
Zeeman coupling $\Delta$ being approximately unchanged.
While our above finite-size data for the height of the Hall conductance
peaks does not seem to allow for a unambiguous extrapolation to the
thermodynamic limit, the suppression of the Hall conductance at large
Zeeman splittings $\Delta\gtrsim0.0075t$ is certainly consistent with
the analytical observations in the infinite system. At large Zeeman couplings,
the singularity in the DOS close to the band edge disappears, and, in turn, the
Hall conductance vanishes.

\begin{figure}
\centering
\includegraphics[scale=0.33]{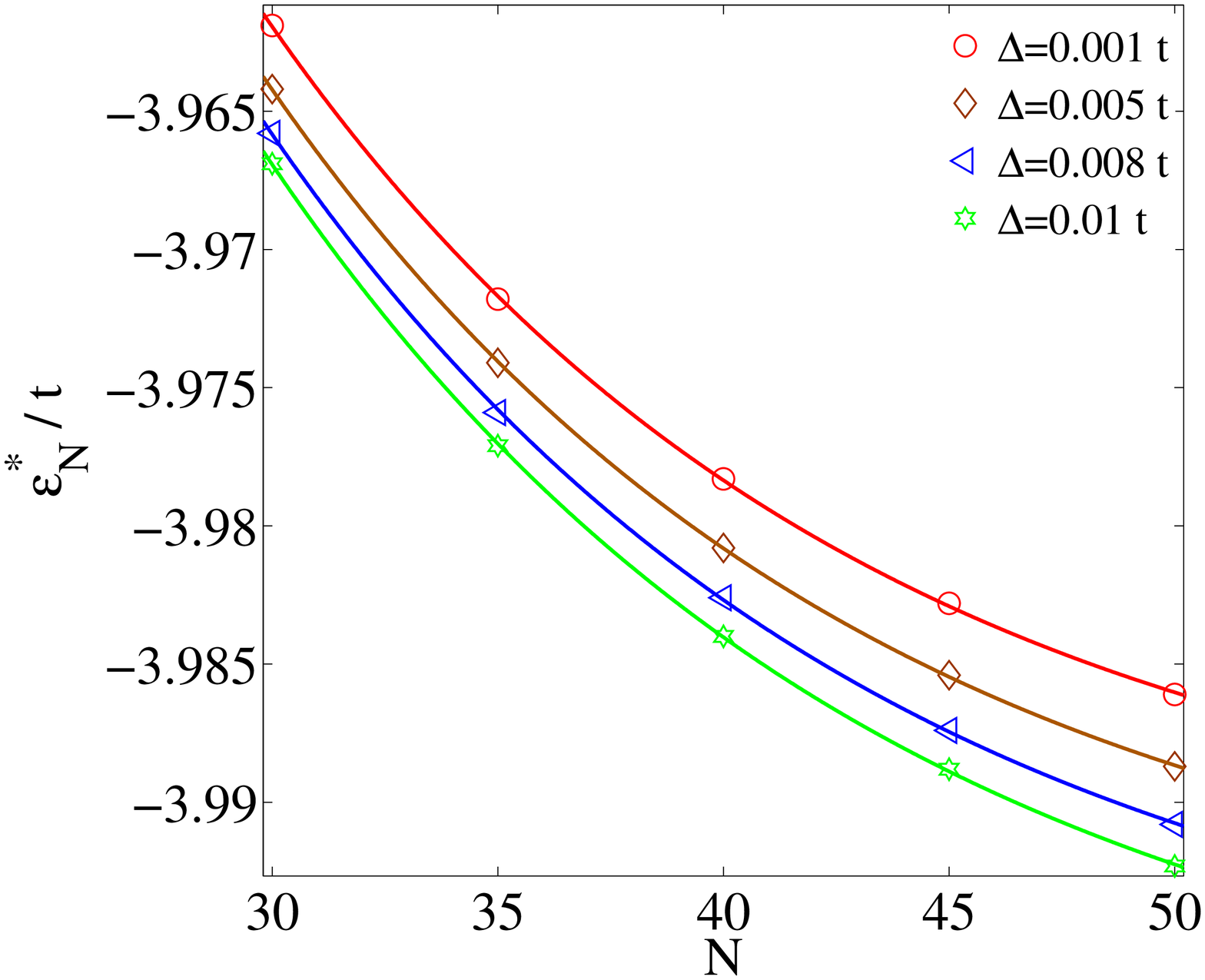}
\caption{(Color online)
Dependence of the position $\varepsilon^{*}_N$
of the Hall conductance peak on the system size for same data sets as
in  Fig.~\ref{CHC_collectiveB001}.
The Rashba parameter is $\lambda=0.01t$. }
\label{pp_vs_ss}
\end{figure}

\begin{figure}
\centering
\includegraphics[scale=0.33]{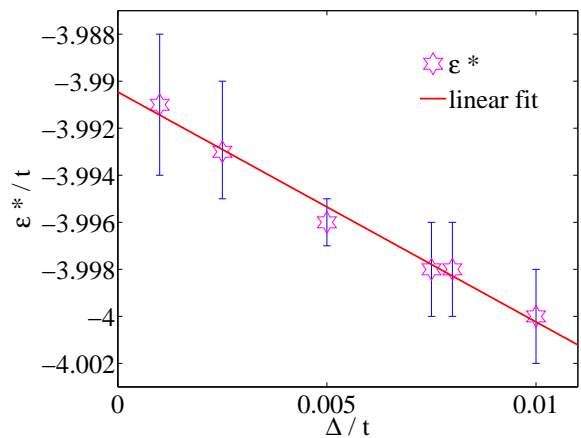}
\caption{Energy position $\varepsilon^*$ of the Hall conductance peak at the
lower band edge extrapolated to an infinite system
versus the magnetic coupling $\Delta$:
the dependence is well fitted by a linear function. }
\label{err_bar}
\end{figure}

Moreover, as also seen in Fig.~\ref{CHC_collectiveB001}, for all
Zeeman couplings,
the position $\varepsilon^{*}_N$
of the peak shifts to lower energies , i.e. towards the bottom of
the band, with increasing system size. Fig.~\ref{pp_vs_ss} shows
the finite-size behavior of the peak position for same six data sets as
in  Fig.~\ref{CHC_collectiveB001}.
All data sets can be smoothly fitted by an exponential function
which allows for an
extrapolation to the limit of an infinite system,
$\varepsilon^{*}=\lim_{N\to\infty}\varepsilon^{*}_N$.
The dependence of
$\varepsilon^{*}$
on the magnetization $\Delta$ is shown in  Fig.~\ref{err_bar}. Clearly,
$\varepsilon^{*}$ linearly decreases with increasing  $\Delta$.
Note, however, that for large Zeeman splitting the Hall conductance peaks
are suppressed with increasing system size although their position can still
be followed as a function of $N$.
On the other hand, even for the smallest Zeeman gap of $\Delta=0.001t$
considered here, the infinite-volume peak position
$\varepsilon^{*}$ lies at an energy where only the lower dispersion branch
of Eq.~(\ref{discdis}) is occupied. This is in accordance with recent
theoretical predictions based on diagrammatic perturbation theory
by Nunner {\em et al.} who concluded that a finite Hall conductance can only
occur at low Fermi energies such that only the lower subband is
occupied \cite{Nunner07,Kato07}.

\subsubsection{In-plane Magnetization}

We now turn to the case of magnetic impurities polarized in the plane
of the 2DEG.
As shown in Fig.~\ref{2DEG}, the Hall current is measured
along the $y$-direction in leads up and down, while the Hall Voltage
is applied along the $x$-direction between leads left and right.
Here we do not observe a Hall current if the in-plane magnetization
is parallel to the voltage since here for charge carriers with wave vector
along the $y$-direction the spin coupling resulting from the
Rashba interaction and the Zeeman coupling are just parallel to each other.
However, a finite Hall current can occur for other in-plane directions of the
magnetization.
\begin{figure}
\centering
\includegraphics[scale=0.35]{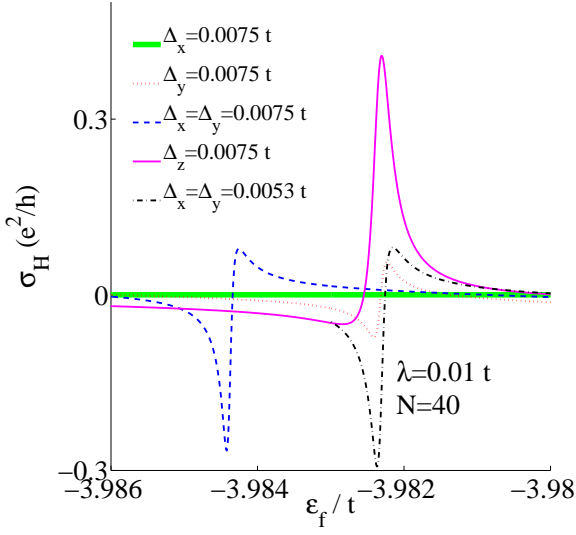}
\caption{(Color online) Hall conductance for a Rashba parameter of
$\lambda=0.01t$ and various directions of magnetization with the non-zero
components of $\vec\Delta$ stated in the legend.
For magnetization along the $x$-direction no Hall current occurs.}
\label{in_plane}
\end{figure}

Fig.~\ref{in_plane} shows the Hall conductance for several directions
of magnetization. For magnetization along the $x$-direction
no Hall current occurs, and for a magnetization pointing in the
$y$-direction with $\Delta_y=0.0075t$
we observe a maximum of the Hall conductance which occurs
at the same energy as for the previous case of strictly out-of-plane
magnetization $\Delta_z=0.0075t$, but is smaller in magnitude.
For a magnetization pointing in the $(1,1,0)$-direction with
 $\Delta_x=\Delta_y=0.0053t$ (fulfilling $|\vec\Delta|=0.0075t$)
we find a conductance peak at the same position in energy but of different
shape. If the magnitude of the Zeeman splitting is increased to
$\Delta_x=\Delta_y=0.0075t$ the peak approximately maintains its shape but
is shifted towards lower energies.
\begin{figure}
\centering
\includegraphics[scale=0.35]{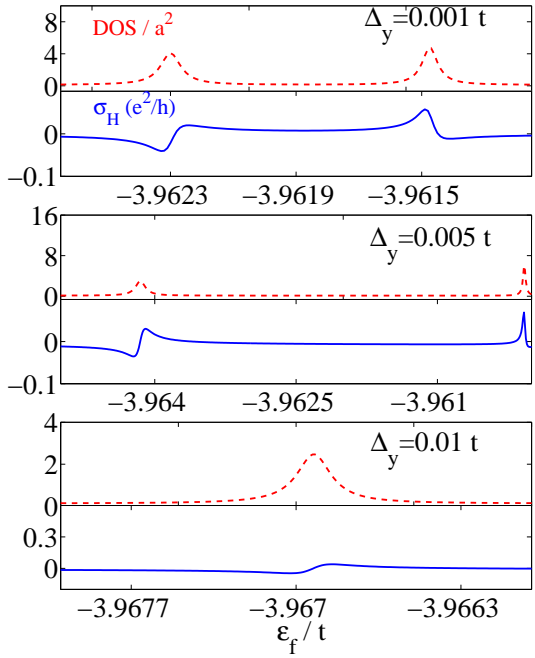}
\caption{(Color online) DOS (red dashed line) in units $1/ta^2$ and
Hall conductance (blue solid line) in units of $e^2/h$
near the lower band edge for an in-plane magnetization polarized along
the $y$-direction, for a system of linear size $N=30$ and for three different
 magnitudes of the Zeeman splitting. This picture is analogous to
Fig.~\ref{DOS_CHC_comp} for the case of a perpendicular magnetization.
Even in the case of an in-plane magnetization we observe the
correspondence between the extrema DOS and Hall
conductance.}
\label{DOS_CHC_y}
\end{figure}
We focus now on an in-plane magnetization totally polarized along the
$y$-direction. Fig.~\ref{DOS_CHC_y} in the ``in-plnae analogue'' of
Fig.~\ref{DOS_CHC_comp} and shows the correspondence between the extrema of the
DOS and those of the Hall conductance for an in-plane magnetization totally
magnetized along the $y$-direction for a system of linear size $N=30$ and for
three different values of the magnitude of the Zeeman splitting. As in the case
of a perpendicular magnetization, see Fig.~\ref{DOS_CHC_comp}, we observe a
correspondence between the extrema of the DOS and the Hall conductance at the
same Fermi energy.
\begin{figure}
\centering
\includegraphics[scale=0.37]{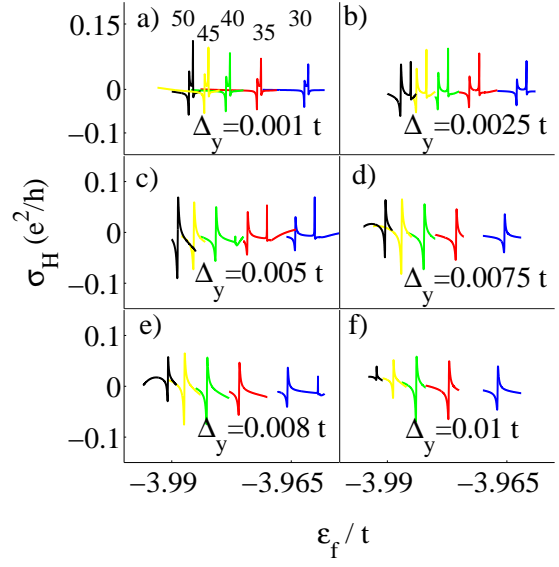}
\caption{(Color online) Hall Conductance near the bottom of the band
for a Rashba parameter of $\lambda=0.01t$ in the case of an in-plane
magnetization along the $y$-direction and for various Zeeman couplings
$\Delta_y$. The linear system size varies from $N=30$ (blue, right), $N=35$ (red),
$N=40$ (green), $N=45$ (yellow), and $N=50$ (black, left). This picture can be
directly compared with Fig.~\ref{CHC_collectiveB001} for the case of a
perpendicular magnetization.}
\label{CHC_By_tot}
\end{figure}

In Fig.~\ref{CHC_By_tot} we show the Hall
conductance for an in-plane magnetization along the $y$-direction
varying between $\Delta_y=0.001 t$ and
$\Delta_y=0.01 t$, and linear system sizes between $N=30$ and $N=50$. The
Rashba parameter is again $\lambda=0.01 t$.
Obviously, the position of Hall peaks shifts to lower energy with increasing
systems size,
analogously as in Fig.~\ref{CHC_collectiveB001} for the case of a perpendicular
magnetization.
Moreover, for the smallest Zeeman coupling considered here, $\Delta_y=0.001 t$,
the height of the peaks clealy grows with systems size, while
for the largest Zeeman coupling of $\Delta_y=0.01 t$ the opposite
behavior is observed. However, we cannot outline any trend for
intermediate values of the Zeeman coupling,
see Fig.~\ref{CHC_By_tot} b), c), d), e).
\begin{figure}
\centering
\includegraphics[scale=0.33]{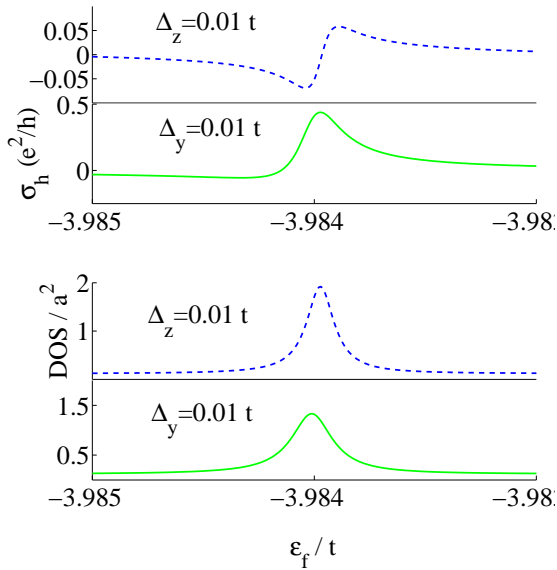}
\caption{(Color online) Hall conductance and DOS for a system of
linear dimension $N=40$ and for a perpendicular magnetization of amplitude
$\Delta_z=0.01 t$ (blue dashed line) and an in-plane magnetization $\Delta_y=0.01 t$ (green solid line)}
\label{DOS_HC_P_I}
\end{figure}
Another interesting finding is that the energetic position of the Hall signal
coincides for in-plane and perpendicular magnetization of the same
magnitude.
This result is shown in Fig.~\ref{DOS_HC_P_I}, where the linear size of the
system is $N=40$ and the Zeeman coupling is chosen to be
$\Delta_z=\Delta_y=0.01 t$. Here, in the two cases of perpendicular and
in-plane magnetization the shape of the DOS and Hall conductance peaks varies
but their maxima exactly coincide .
The dependence of the Hall conductance
on direction and magnitude of the in-plane magnetization is to
be explored further in the future.

\subsection{Magnetically disordered systems}

So far we have studied homogeneously magnetized systems with
each lattice site carrying a magnetic ion whose spin provides a
Zeeman field for the carrier spins. We now consider the case where
only a given fraction $\nu$ of lattice sites is occupied by a magnetic ion.
This scenario accounts for the situation in real ferromagnetic semiconductor
nanostructures. We will concentrate again on magnetizations along the
$z$-direction perpendicular to the plane of the 2DEG. To be specific, we choose
at random a given fraction of lattice sites to be occupied with a magnetic ion
and average our results for the Hall conductivity over typically 20
of such disorder realizations, which, by inspection of the data, turns out
to be sufficient. In order to facilitate the comparison with
our previous results for magnetically homogeneous systems we adjust the
magnitude of each local coupling $\Delta_{m,n}$ such that the average Zeeman
coupling $\Delta:=(\sum_{m,n}\Delta_{m,n})/N^2$ is constant, i.e.
$\Delta_{m,n}=\Delta/\nu$.
\begin{figure}
\centering
\includegraphics[scale=0.35]{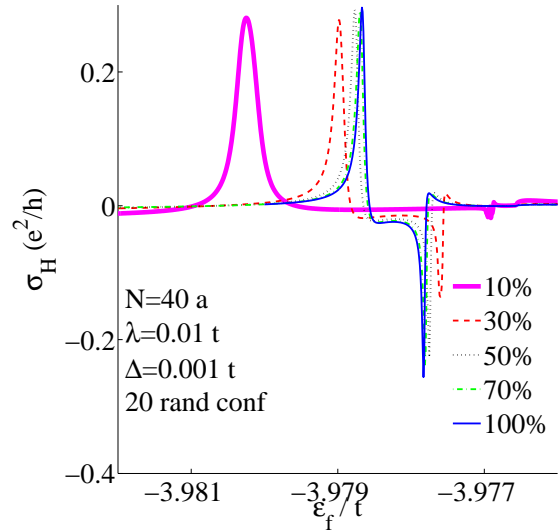}
\caption{(Color online) Hall conductance close to the lower band edge
for different fractions of magnetically occupied sites for a Rashba parameter
of $\lambda=0.01t$ and an average magnetization of $\Delta=0.001 t$. The
linear size of the system is N=40.}
\label{AVG40_tot}
\end{figure}
Fig.~\ref{AVG40_tot} shows the Hall conductance
for different fractions of magnetically occupied sites for a Rashba parameter
of $\lambda=0.01t$ and an average magnetization of $\Delta=0.001 t$.
The data is averaged over 20 random disorder configurations.
\begin{figure}
\centering
\includegraphics[scale=0.37]{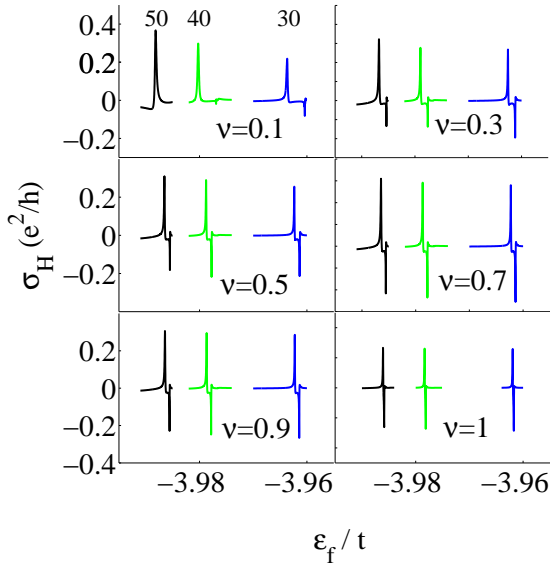}
\caption{(Color online) Hall conductance maxima for systems of linear dimension $N=30$ (blue, right), $N=40$ (green), $N=50$ (black, left) and for several fraction of magnetically occupied sites.}
\label{peaks}
\end{figure}
Fig.~\ref{peaks} shows the dependence of the Hall conductance both on the
fraction of magnetized sites as well as on system size.
Again, the energetic position of the hall maxima moves towards lower energies
with increasing system size, cf. Figs.~\ref{CHC_collectiveB001},\ref{CHC_By_tot}.
Moreover, for fractions  $\nu=0.3\dots 0.9$ of magnetized sites, the position of
the Hall peaks is very close to that of the uniform magnetization, $\nu=1$, and
the height of peaks is approximately constant in all cases. This is different
from the smallest fraction considered here, $\nu=0.1$, where height increases
with system size, and also the energetic positions differ from the other cases.

\section{Conclusions and Outlook}

\label{concl}
Anomalous Hall transport is still the subject of a long-standing
theoretical discussion. In the present paper
we have report on the, to our knowledge, first investigation of this
phenomenon using the Landauer-B\"uttiker formalism.
Specifically, we have studied
a two-dimensional electron gas with Rashba spin-orbit coupling and a
magnetization provided by localized magnetic moments.
Our system is
described by a discretized tight-binding model in a four-terminal geometry.
We have considered both the case of  ballistic transport in
disorder-free systems with homogeneous magnetization
as well as magnetically disordered systems. In the former case we have also
distinguished between different directions of the magnetization. In particular,
a magnetization lying entirely in the plane of the 2DEG can also be
interpreted as a genuine magnetic field which couples, in a strictly
two-dimensional system, only to the spin of the electron but not to its
orbital degrees of freedom.

In particular, we have demonstrated numerically
a close connection between singularities in the density of states
and peaks in the Hall conductance close to the lower band edge. Consistent with
previous theoretical studies based on diagrammatic perturbation
expansions, these peaks occur at Fermi energies where only the
lower dispersion branch is occupied \cite{Nunner07,Kato07}.
Moreover, for large magnetization
the Hall conductance is, along with the density of states, suppressed.
This numerical finding can be understood from analytical properties of
the underlying model in the limit of an infinite system.

Future investigation will include a more detailed understanding of anomalous
Hall transport in the presence of an in-plane magnetization, and
the effects of magnetic disorder.

\acknowledgments{This work was supported by the SFB 689
``Spin Phenomena in reduced Dimensions''.}

\appendix
\section{Green큦 function of a semi-infinite lead}
\label{greenfunct}
Here we give some more technical details regarding the calculation of the
Green큦 function of Eq.(\ref{green}) for a semi-infinite lead.

For a semi-infinite non-interacting lead with hard wall confinement and
a constant width $L$, the transverse wave functions are
\begin{equation}
\tilde{\chi}_m(y)=\sqrt{\frac{2}{L}}\sin\left(\frac{m \pi y}{L}\right)
\end{equation}
or
\begin{equation}
 \chi_ m(y_j)=\sqrt{\frac{2}{M+1}}\sin\left(\frac{m \pi j}{M+1}\right),
\end{equation}

where $y_i=ja$ and $M$ is the number of sites in the transverse
direction, such that $L=(M+1)a$. The longitudinal lattice wave functions
are
\begin{equation}
 \phi_k(x)=\sqrt{\frac{2}{L}}\sin ( k x ),
\end{equation}
which, by substituting $x=a$, which means that we are considering points
$x$ at the first slice of the semi-infinite lead, transforms into
\begin{equation}
 \phi_k(x=a)=\sqrt{\frac{2 a}{L}}\sin ( k a ).
\end{equation}

Finally, at the first slice $x=a$ the total wave function reads
\begin{equation}
\psi_{m,k}=\sqrt{\frac{2a}{L}}\chi_m(y)\sin ( k a ),
\end{equation}
which is an eigenfunction of the Hamiltonian $H_0
\psi_{m,k}=E_{m,k}\psi_{m,k}$, where the dispersion relation is given by
\begin{equation}
 E_{m,k}=2 t\left(1-\cos \frac{m \pi}{M+1}\right)+2 t (1-\cos (k a)).
\end{equation}

 Inserting the wave function in the eigenfunctions expression of the
Green큦 function, see \cite{datta}, the Green큦 function at the first
slice $x=a$ is
\begin{equation}
 G^r(a,y_i;a,y_j)=\frac{2 a}{L}\sum_{m,k
>0}\frac{\chi_{m}(y_i)\chi_{m}^*(y_j)\sin^2(k a)}{E-E_{m,k}+i \eta}.
\end{equation}
In the limit $L\rightarrow\infty$, we may replace the sum over $k$ by
the integral $(L/\pi)\int_0^{\pi/a}dk$ and substitute $k a=\theta$ to
obtain
\begin{equation}\label{int}
\begin{array}{ll}
G^r(a,y_i;a,y_j)=&1/\pi t\sum_{m}\chi_{m}(y_i)\chi_{m}^*(y_j)\\
&\int_0^{\pi} d \theta \sin^2\theta/(Q+\cos\theta+i \eta),
\end{array}
\end{equation}
where we made the following replacement
\begin{equation}
 Q=\frac{E}{2 t}-2+\cos\frac{m \pi}{M+1}.
\end{equation}
In Eq.(\ref{int}) the integrand function $\sin^2\theta/(Q+\cos\theta+i
\eta)$ is an even function of $\theta$, therefore the integral can be
written as a symmetric integral from $-\pi$ to $\pi$. By writing in the
integral the sine and cosine functions in terms of their exponential
form and performing the substitution $z=\exp(i\theta)$, the integral
turns into a closed contour integral along the unit circle in the
complex plane, allowing us to write the Green function as
\begin{equation}
\begin{array}{ll}
G^r(a,y_i;a,y_j)&=1/(2\pi i t)\sum_{m}\chi_{m}(y_i)\chi_{m}^*(y_j)\\
&\int_{C} d z (1-z^2)/(z^2+2 z Q+1),
\end{array}
\end{equation}
where $\int_{C}$ stands for the integration on a closed circuit of
radius $\mid z\mid=1$.
By solving the integral with the use of the theorem of residues, we
obtain
\begin{equation}\label{green2}
\begin{array}{ll}
G^r(a,y_i;a,y_j)&=1/(2\pi i t)\sum_{m}\chi_{m}(y_i)\chi_{m}^*(y_j)\\
& 2\pi i{\mbox R}_{z_0} (1-z^2)/(z^2+2 z Q+1),
\end{array}
\end{equation}
where ${\mbox R}_{z_0}$ indicates the residual calculated at the pole
$z=z_0$ which depends on $Q$, i.e.
\begin{equation}
z_0=\left\{
 \begin{array}{ll}
 -Q+\sqrt{Q^2-1} & \mbox{for}\;\;\;\; Q>1\\
 -Q-\sqrt{Q^2-1} & \mbox{for}\;\;\;\; Q<-1\\
 -Q+i\sqrt{1-Q^2} & \mbox{for}\;\;\;\; \mid Q\mid\leq 1\\
\end{array}\right.
\end{equation}
Inserting the pole in Eq.(\ref{green2}), we obtain the final expression
for the green function
\begin{equation}\label{green2p}
 G^r(a,y_i;a,y_j)=\frac{1}{t}\sum_{m}\chi_{m}(y_i)\chi_{m}^*(y_j)F(Q),
\end{equation}
with
\begin{equation}
F(Q)=\left\{
 \begin{array}{ll}
 Q-\sqrt{Q^2-1} & \mbox{for}\;\;\;\; Q>1\\
 Q+\sqrt{Q^2-1} & \mbox{for}\;\;\;\; Q<-1\\
 Q-i\sqrt{1-Q^2} & \mbox{for}\;\;\;\; \mid Q\mid\leq 1\\
\end{array}\right.
\end{equation}

\end{document}